\documentclass[column,floatfix,superscriptaddress,showpacs,showkeys,nofootinbib,preprint]{revtex4}%
\usepackage{epsfig}
\usepackage{amssymb,latexsym,amsmath}
\newcommand{\eq}[1]{\begin{align} #1 \end{align}}

\usepackage{amsmath}
\usepackage{graphicx}

\newcommand{\ce}[2]{\mbox{#1--#2\%}}

\begin{document}

\title{Bose-Einstein condensation of pions in heavy-ion collisions at the CERN Large Hadron Collider (LHC) energies}

\author{Viktor Begun}
 \affiliation{Institute of Physics, Jan Kochanowski University, PL-25406~Kielce, Poland}
 \affiliation{Bogolyubov Institute for Theoretical Physics, 03680 Kiev, Ukraine}

\author{Wojciech Florkowski}
 \affiliation{The H. Niewodnicza\'nski Institute of Nuclear Physics, Polish Academy of Sciences, PL-31342 Krak\'ow, Poland}

\date{\today}

\begin{abstract}
We analyse in detail the possibility of Bose-Einstein condensation
of pions produced in heavy-ion collisions at the beam energy
$\sqrt{s_{\rm NN}}$ = 2.76 TeV. Our approach is based on the
chemical non-equilibrium thermal model of hadron production which
has been generalised to include separately the contribution from
the local zero-momentum state. In order to study both the hadronic
multiplicities and the transverse-momentum spectra, we use the
Cracow freeze-out model which parameterises the flow and
space-time geometry of the system at freeze-out in a very economic
way. Our analysis indicates that about 5\% of all pions may form
the Bose-Einstein condensate.
\end{abstract}

\pacs{25.75.-q, 25.75.Dw, 25.75.Ld}

\keywords{relativistic heavy-ion collisions, thermal models of
hadron production, Bose-Einstein condensation, Large Hadron Collider (LHC)}

\maketitle

%
\section{Introduction}
\label{sect:Intro}
%

Thermal models of hadron production serve us as the basic tool to
obtain general information about the properties of matter produced
in heavy-ion
collisions~\cite{Koch:1985hk,Cleymans:1992zc,Schnedermann:1993ws,Sollfrank:1993wn,Braun-Munzinger:1994xr,
Becattini:1995if,Braun-Munzinger:1995bp,Cleymans:1996cd,Becattini:1997uf,Becattini:1997rv,
Yen:1998pa,Cleymans:1998fq,Gazdzicki:1998vd,Letessier:1998sz,Cleymans:1999st,Braun-Munzinger:1999qy,Becattini:2000jw,
Braun-Munzinger:2001ip,Florkowski:2001fp,Broniowski:2001we,Broniowski:2001uk,Broniowski:2002wp,
Retiere:2003kf,BraunMunzinger:2003zd,Becattini:2003wp}. The
successes of such models in description of different collision
processes indicate typically that the produced matter is well
thermalised and exhibits collective behavior. This, in turn,
suggests the formation of the equilibrated system of quarks and
gluons at earlier stages of the collisions.

The newest LHC data on hadron production indicates that the most
common thermal approach, based on the chemical and thermal
equilibration assumption, fails to describe the ratio of protons
to pions~\cite{Abelev:2013vea}. The mean multiplicities of protons
and anti-protons also deviate from the fit for about three
standard deviations~\cite{Stachel:2013zma}. In addition, most of
the hydrodynamic calculations, that have been successfully used to
reproduce the harmonic flow coefficients, have problems with the
correct predictions for very low transverse-momentum spectra of
pions~\cite{Gale:2012rq,Abelev:2012wca,Abelev:2013vea,Molnar:2014zha}.

In our previous works~\cite{Begun:2013nga,Begun:2014rsa} we have
demonstrated that these two problems may be solved by the
assumption that matter produced in heavy-ion collisions at the LHC
energies is formed out of chemical
equilibrium~\cite{Koch:1985hk,Sollfrank:1993wn,Letessier:1998sz}.
One of the predictions of the chemical non-equilibrium model is
that the pion abundances are characterized by the non-zero value
of the chemical potential\footnote{The equilibrium
model~\cite{Melo:2015wpa} also shows the hints of pion chemical
potential at the LHC.} which is very close to the critical value
for the Bose-Einstein condensation (BEC). This suggests a possible
onset of BEC at the LHC energies.

The recent analysis of two- and three-pion correlations done by
ALICE collaboration~\cite{Abelev:2013pqa} shows an intriguing
result that a coherent fraction in charged pion emission may reach
23$\%$. The main aim of our present work is to determine
quantitatively the amount of particles in condensate at different
centralities using mean multiplicities and spectra of the
particles.

The fact that the pion chemical potential is very close to its
critical value requires that thermal analyses in this case should
be performed in a more careful way with the explicit treatment of
the local zero-momentum state \cite{Begun:2014aha}. As the
inclusion of the pion ground state effects is relatively easy in
the analyses of the ratios of hadronic abundances, a fully
covariant framework describing local formation of the condensate
in an expanding medium is not at hand. The formulation of such a
framework requires that the Cooper-Frye formula is generalized in
a very special way. In this work we demonstrate how this can be
achieved for boost-invariant and cylindrically symmetric systems.
We make the corresponding changes in the previously used by us
SHARE~\cite{Petran:2013dva} and
THERMINATOR~\cite{Kisiel:2005hn,Chojnacki:2011hb} codes.

The hadronic data are analyzed within three possible scenarios:
using the full equilibrium model (EQ), the chemical
non-equilibrium model (NEQ) that have been used
in~\cite{Begun:2013nga,Begun:2014rsa}, and the newly developed
chemical non-equilibrium model with Bose-Einstein condensate~
(BEC)~\footnote{Note that we use the same acronym to denote both
the Bose-Einstein condensation and our framework which explicitly
includes the condensate formation.}. We include all available LHC
data on Pb+Pb collisions at $\sqrt{s_{\rm
NN}}=2.76$~TeV~\cite{Abelev:2013vea,Abelev:2013xaa,ABELEV:2013zaa,Abelev:2014uua}
in our study using the modified SHARE and use the modified
THERMINATOR in order to check the effect of BEC on the pion
spectra at low
transverse-momenta~\cite{Abelev:2013vea,Abelev:2014ypa}. We also
discuss in greater detail the effects which the data on neutral
pion production may have on our results. Our comparisons between
different thermal frameworks indicate the plausibility of pion
condensation in heavy-ion collisions at the LHC.

The paper is organized as follows: in Section~\ref{sect:BEC} we
describe the role of the ground state and the difference between
EQ, NEQ and BEC models. In Section~\ref{sect:fitstrateg} we
explain the method of our analysis. Sections~\ref{sect:Ratios}
and~\ref{sect:BEC-THERMINATOR} show the results of the analysis of
mean multiplicities and particle spectra, correspondingly.
Section~\ref{sect:CONCLUSIONS} concludes the paper.

%
\section{explicit treatment of hadronic ground states}
\label{sect:BEC}
%

\subsection{Ground state contribution}

Calculating particle multiplicities in a non-interacting quantum
gas, one changes the summation over discrete
momentum levels by the integration over the continuous spectrum
\begin{eqnarray}
 N  &=& \sum_n \frac{g_n}{\exp\left(\frac{\sqrt{p_n^2+m^2}-\mu}{T}\right)\mp1}
 ~\simeq  \int \frac{d^3x\,d^3p}{h^3}\, \frac{g}{\exp\left(\frac{\sqrt{{\bf p}^2+m^2}-\mu}{T}\right)\mp1}
  \nonumber \\
    &=& V\int_0^{\infty} \frac{d^3p}{(2\pi)^3}\, \frac{g}{\exp\left(\frac{\sqrt{{\bf p}^2+m^2}-\mu}{T}\right)\mp1}~
\label{V1}.
\end{eqnarray}
Here $g_n$ is the degeneracy of the $p_n$-th momentum state, $T$
is the system temperature, $m$ is the mass of particles, and $\mu$ is
the chemical potential. The integral over space coordinates gives the volume
of the system $V$, and the factor $(2\pi)^3$ appears since we use the natural
units where $h\equiv 2\pi\hbar$ and $\hbar=1$. For simplicity, we do not specify the
physical character of the chemical potential $\mu$ at the moment and
consider only one type of particles.

If the  chemical potential approaches the value of the particle mass, the
integrand in the second line of Eq.~(\ref{V1}) stays constant for ${\bf p}\rightarrow 0$,
because the singularity of the denominator is cancelled by the integration measure
$d^3p$. However, the first term in the sum over
quantum levels becomes infinite in the limit  $\mu \to m$ at ${\bf p}=0$
 \begin{equation}
 N_{\rm cond} ~=~ \frac{g_0}{\exp\left(\frac{m-\mu}{T}\right)-1}~\rightarrow~\infty~\quad \hbox{for} \quad
 \mu~\rightarrow~m~.\label{N0}
 \end{equation}
Therefore, at the onset of BEC, when $\mu\rightarrow m$, one should
keep the summation over the low momentum states and start the
integration in (\ref{V1}) at $|{\bf p}|>0$. One can show, however, that in the thermodynamic
limit, $V\rightarrow\infty$, one may separate the ${\bf p}=0$
term only and start the integration from zero~\cite{Begun:2008hq}, namely
\begin{eqnarray} \label{replace3}
 N ~&\simeq&~   \frac{g}{\exp\left(\frac{m-\mu}{T}\right)-1}
 ~+~  V\int_0^{\infty} \frac{d^3p}{(2\pi)^3}\, \frac{g}{\exp\left(\frac{\sqrt{{\bf p}^2+m^2}-\mu}{T}\right)-1}
 \nonumber \\
 ~&=&~ N_{\rm cond} ~+~ N_{\rm norm}~= N_{\rm cond} ~+~V \, n_{\rm norm}.
\end{eqnarray}
The degeneracy of the zero level is the same as that of the continuous
spectrum, $g_0=g=2s+1$, where $s$ is the particle spin.

\subsection{Chemical potentials in a hadron gas}

In the case of a hadron-resonance gas the above simple
picture becomes more complicated  --- one should include: the sum
over all existing hadron states, the integral over the mass spectrum of broad resonances,
and the decays of resonances defined by their branching ratios.
In addition, the chemical
potential is different for each particle type $i$. In the case of the
EQ model, the chemical potential becomes a linear combination of the
particle's electric, $Q_i$, baryon,
$B_i$, and strange, $S_i$, charges
 \eq{
 \mu_i^{\rm EQ}~=~Q_i\mu_Q~+~B_i\mu_B~+~S_i\mu_S~.
 }
The chemical potentials $\mu_Q$, $\mu_B$ and $\mu_S$ are the same
in the whole system. They control the conservation of electric
charge, baryon number, and strangeness. In the midrapidity region
at the LHC these quantities are negligible, so it is reasonable to
assume that
 \eq{
 \mu_i^{\rm EQ}~\simeq~0~.
 }
The charge conservation is connected with the conservation of the
{\it difference} of the quark and antiquark numbers. However, a
fast expansion and cooling of the quark-gluon plasma may lead to
the effective conservation of the {\it sum} of quarks and
antiquarks~\cite{Sollfrank:1993wn}. It can happen if the
hadronization is so fast, that the quarks do not have enough time
to annihilate and lower their abundances to the equilibrium values
at the chemical freeze-out temperature. This leads to the
appearance of additional non-equilibrium chemical potential
 \eq{
 \mu_i^{\rm NEQ}~=~(N_q^i+N_{\bar{q}}^i)~\mu_q~+~(N_s^i+N_{\bar{s}}^i)~\mu_s~,
 }
where $N_q^i$ and $N_{\bar{q}}^i$ are numbers of light ($u, d$)
quarks and antiquarks, while $N_s^i$ and $N_{\bar{s}}^i$ are the
numbers of strange quarks and antiquarks in the $i$th hadron. For example, the non-equilibrium
chemical potentials for pions, protons,
kaons, and $\Lambda$'s read:
\eq{
 \mu_{\pi} ~=~ 2\mu_q~,&&\mu_{p} ~=~ 3\mu_q~,&&\mu_{K} ~=~ \mu_q~+~\mu_s~,
 &&\mu_{\Lambda} ~=~ \mu_q~+~2\mu_s~. }
We neglect the contributions from heavier quarks, because they are
suppressed by the Boltzmann factor.

The non-equilibrium chemical potentials may be encoded in the fugacity factors of the form
 \eq{\label{pre-factor}
 \Upsilon_i ~\equiv~ \exp\left(\frac{\mu_i^{\rm NEQ}}{T}\right)
 ~=~\gamma_q^{N_q^i+N_{\bar{q}}^i}\gamma_s^{N_s^i+N_{\bar{s}}^i}.
 }
As a matter of fact, the non-equilibrium hypothesis of hadron
production at freeze-out was originally proposed in terms of
$\Upsilon_i$'s~\cite{Koch:1985hk}, therefore, we will follow the
tradition and use $\gamma_q$ and $\gamma_s$ instead of $\mu_q$ and
$\mu_s$. We have introduced the condensation term $N_{\rm cond}$
given by Eq.~(\ref{replace3}) in the latest version of
SHARE~\cite{Petran:2013dva}. The calculations done in this way are
denoted below as the BEC case.

%
\section{Fitting strategy}
\label{sect:fitstrateg}
%

\subsection{Centrality selection}

There are several technical issues that we have to address now. Different particles
are measured in different centrality bins. This leads to problems with the interpretation
of the data. The authors of Ref.~\cite{Becattini:2014hla} solved
this problem by redistributing heavy strange particles in smaller
centrality bins, in which light particles such as pions are
measured. However, we have found that this strategy introduces an
uncertainty in the thermal analyses, which leads to large
errors for the output parameters, especially for $\gamma_q$ and $\gamma_s$.

Therefore, herein we do the opposite and merge the particles
measured with the finer centrality steps into larger centrality
sets at which strange particles are measured. This gives us six
centrality intervals: \ce{0}{10}, \ce{10}{20}, \ce{20}{40},
\ce{40}{60}, \ce{60}{80}, and \ce{80}{90}. The corresponding
errors are recalculated following the standard procedure of error
propagation.

\subsection{Multiplicities vs. ratios}

The experiment provides both multiplicities (strictly speaking the rapidity densities)
and particle ratios. Some authors prefer to fit particle ratios instead of
multiplicities, in order to have fewer number of parameters and to
get rid of the system's volume. However, one should be careful
with this procedure, especially for the non-equilibrium models,
because with given $n$ multiplicities one may construct $n
(n-1)/2$ ratios. This gives quite substantial number of ratios for
large $n$, which are correlated.

In practice, one usually
restricts oneself to a smaller and experimentally available set of
ratios, and thus studies only a sub-domain of the whole system.
This strategy affects the results, in particular for the NEQ
models, where each particle has its own pre-factor
$\Upsilon_i$~(\ref{pre-factor}).
For example, the multiplicity of protons is approximately proportional to
$\Upsilon_{p}=\gamma_q^3$, while that of  pions to
$\Upsilon_{\pi}=\gamma_q^2$. Therefore the ratio of protons to
pions will give us the factor $\gamma_q$, which is much less
sensitive to the change of $\gamma_q$ than the mean
multiplicities of protons and pions. The ratio of lambda to kaon
cancels the information on strangeness,
$\Upsilon_{\Lambda}/\Upsilon_K=(\gamma_q^2\gamma_s)/(\gamma_q\gamma_s)=\gamma_q$,
and gives us as much information as proton to pion ratio, etc.
Therefore, even if one takes all $n (n-1)/2$ ratios, one may get
different temperatures and other parameters for the same set of
$n$ particles.

Another important aspect of the analysis is that ALICE measures only the rapidity
densities. Below, we argue that Eq. (\ref{replace3}) can be transformed in this case to the form
\begin{eqnarray}
\frac{\Delta N}{ \Delta y } ~
= \frac{dV}{dy} \, n_{\rm norm} + N_{\rm cond},
\label{master}
\end{eqnarray}
where $dV/dy$ is the volume of the produced particles per unit rapidity.
Note that the volume does not cancel in the ratios of the rapidity densities
in the case with the condensate included.

\subsection{Neutral pion effects on the fits}

The BEC model  gives
typically a very good fit for all observed particles. However, the
calculated number of the not yet measured $\pi^0$ mesons is
extremely large. This is so, because the fit suggests a high value
of $\gamma_q$, at which the neutral pions, as the lightest
particles, condense first. This behaviour suggests that one should
make a reasonable estimate of the $\pi^0$ multiplicity and use it
as an input in the BEC fit.

The isospin symmetry of strong interactions suggests that  the
number of $\pi^0$'s must be the same as the number of positively
or negatively charged pions. Although the
isospin symmetry is in fact slightly broken (in particular, $\pi^0$ is
lighter than $\pi^{\pm}$) we neglect this effect in
order to get a lower bound for the amount of condensate, and use
the constraint
 \eq{\label{Pi0}
 N_{\pi^0}(c) ~=~\frac{1}{2}~\left(N_{\pi^+}(c)+N_{\pi^-}(c) \right).
 }
Equation~(\ref{Pi0}) represents an estimate of a
possible production rate. The error bars are recalculated as
follows
 \eq{\label{Pi0Err}
 \Delta N_{\pi^0}(c) ~=~\frac{1}{2}~\left(\Delta N_{\pi^+}(c)+\Delta N_{\pi^-}(c) \right).
 }
Since the experimental errors for the $\pi^+$ and $\pi^-$ mean
multiplicities are practically the same, we obtain almost the same
error bars for $\pi^0$. We note that it is much more difficult to
measure $\pi^0$'s than charged pions. At present, the spectrum for
$p_T>700$~MeV is known only. It gives just about $1/3$ of all
expected $\pi^0$'s. In the $p_T$ range where both $\pi^{\pm}$ and
$\pi^0$ are measured, the latter has about twice bigger relative
errors. The error bars for $\pi^0$-s also increase fast, while
going from high to low $p_T$ in the existing
data~\cite{Abelev:2014ypa}. Therefore, we think that
(\ref{Pi0Err}) is a reasonable estimate.


\section{Fit of hadronic multiplicities}\label{sect:Ratios}

We perform our analysis using the rapidity densities of
$\pi^{\pm}$, $p$, $\bar{p}$, $K^{\pm}$, $K_S^0$, $\Lambda$,
$\phi$, $\Xi^{\pm}$ and $\Omega^{\pm}$. In order to include the
effect of the neutral pions on the fits, we also include the
estimate based on  Eqs.~(\ref{Pi0})~and~(\ref{Pi0Err}).
This gives us 14 data points for the centralities \ce{0}{10},
\ce{10}{20}, \ce{20}{40}, \ce{40}{60}, \ce{60}{80}, and 10 data
points for the centrality bin \ce{80}{90},
where $\Xi^{\pm}$ and $\Omega^{\pm}$
are not measured. We do not include the short-living
$K^*(892)^0$~\cite{Abelev:2014uua} ---
in this way we may check if it follows
the pattern obtained for other long-living particles.
By doing so we further examine the applicability of the single
freeze-out concept, see our discussion in~\cite{Begun:2014rsa}.

\begin{figure}[h!]
 \epsfig{file=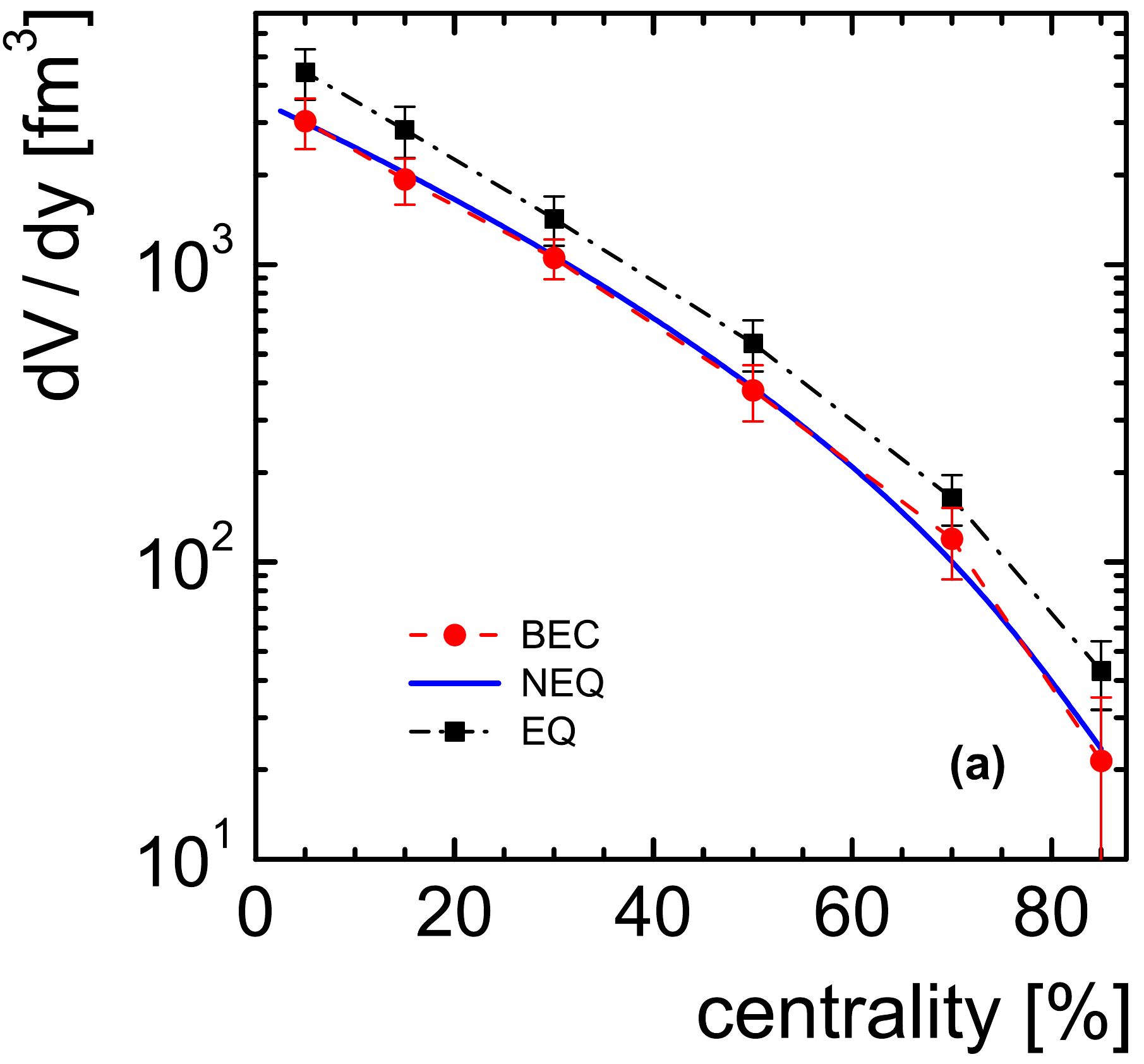,         width=0.49\textwidth}\;    
 \epsfig{file=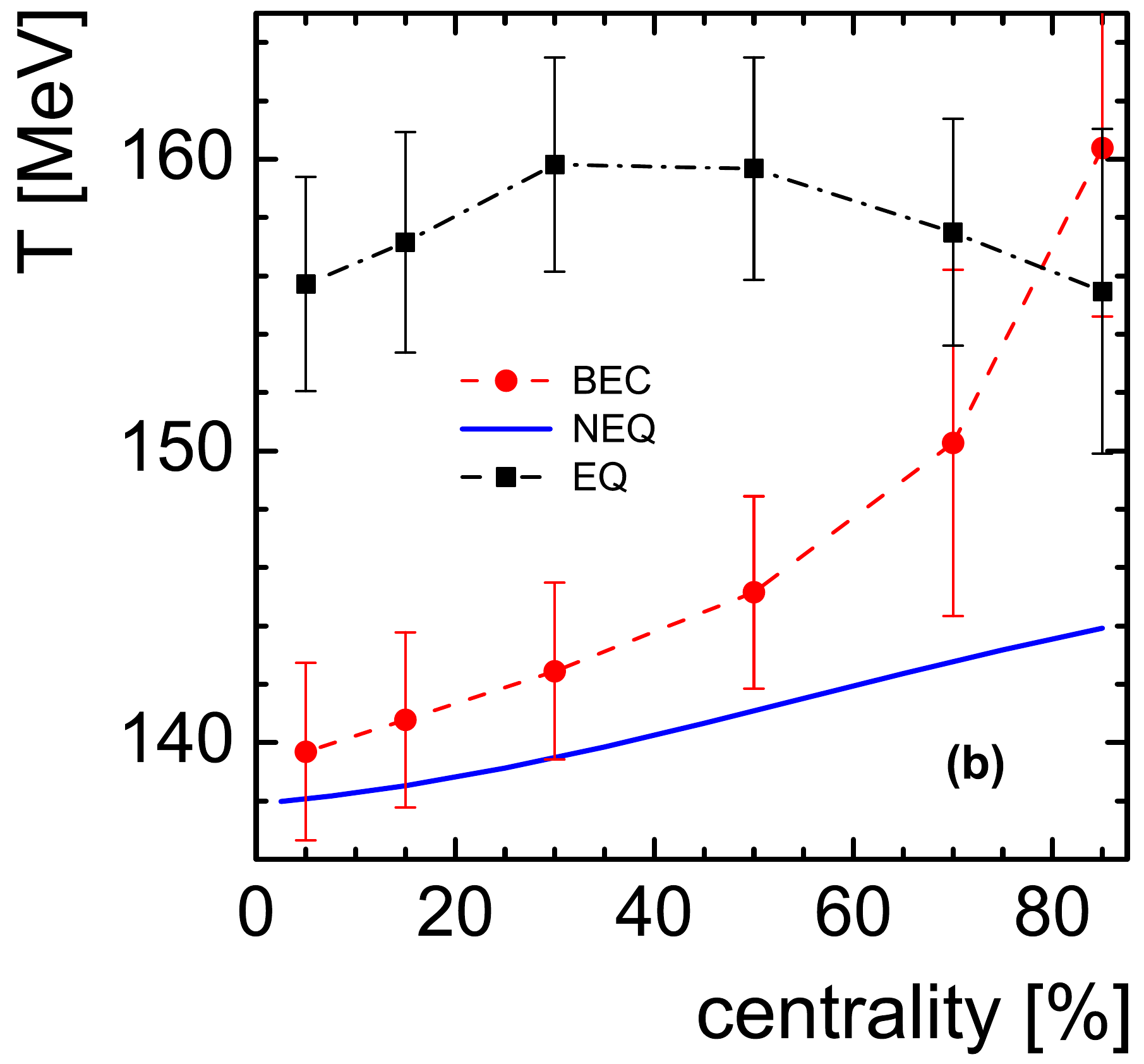,         width=0.49\textwidth}      
 \caption{(Color online) The left figure shows the system volume, while the right figure shows the system temperature obtained as the fit to mean multiplicities of hadrons,
 which are produced in PbPb collisions at the beam energy $\sqrt{s_{\rm NN}}=$2.76 TeV. The BEC lines correspond to the non-equilibrium model {\it with} the Bose condensate
 in the ground state~(\ref{replace3}). The NEQ lines correspond to the non-equilibrium model {\it without} the Bose condensate, and the EQ lines are obtained for the equilibrium
 model. The NEQ lines are from the paper~\cite{Begun:2014rsa}, while the EQ and BEC lines are from~\cite{Begun:2014aha}.
 }\label{fig:V-T}
\end{figure}

\begin{figure}[h!]
 \epsfig{file=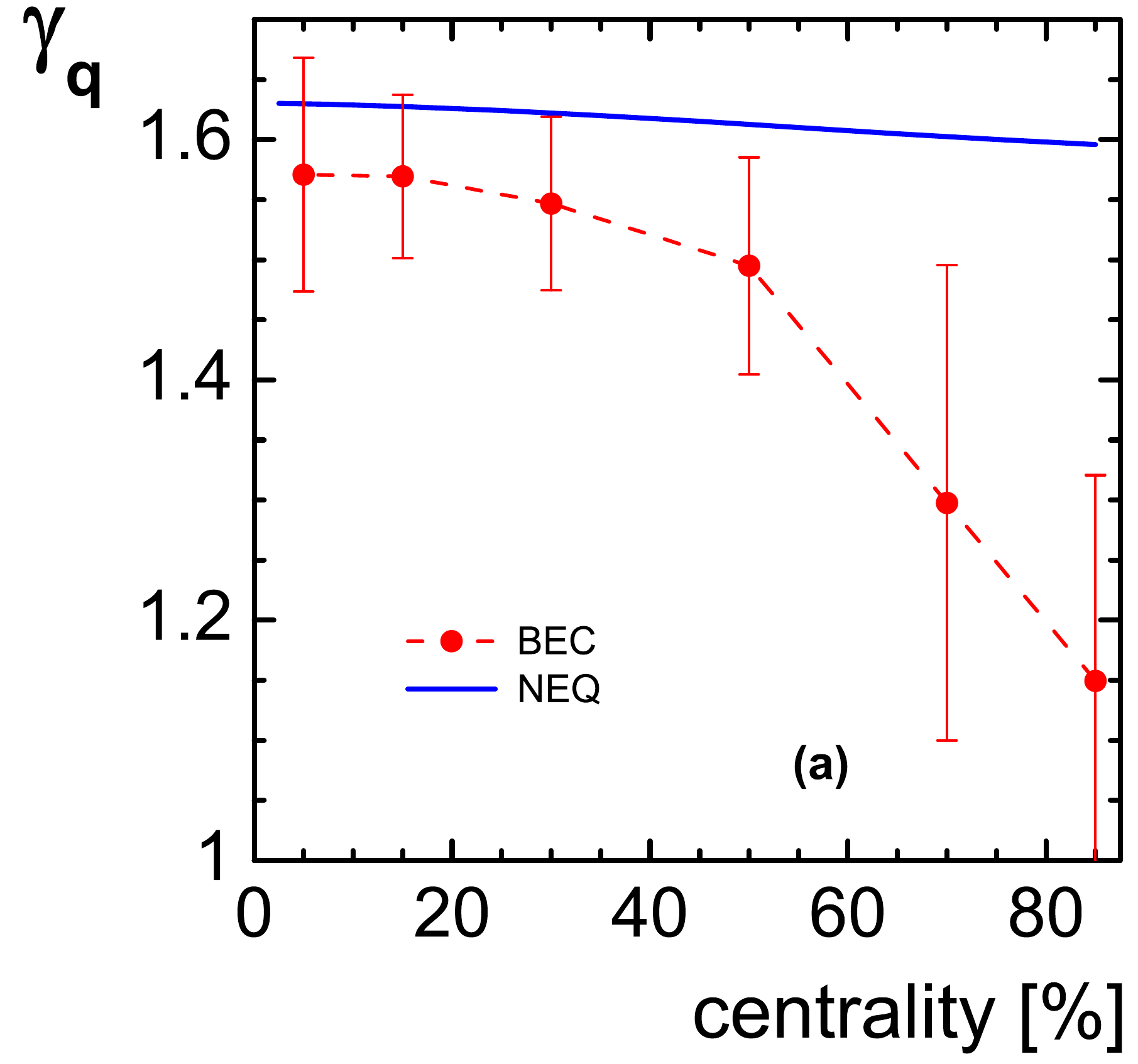,    width=0.49\textwidth}\;     
 \epsfig{file=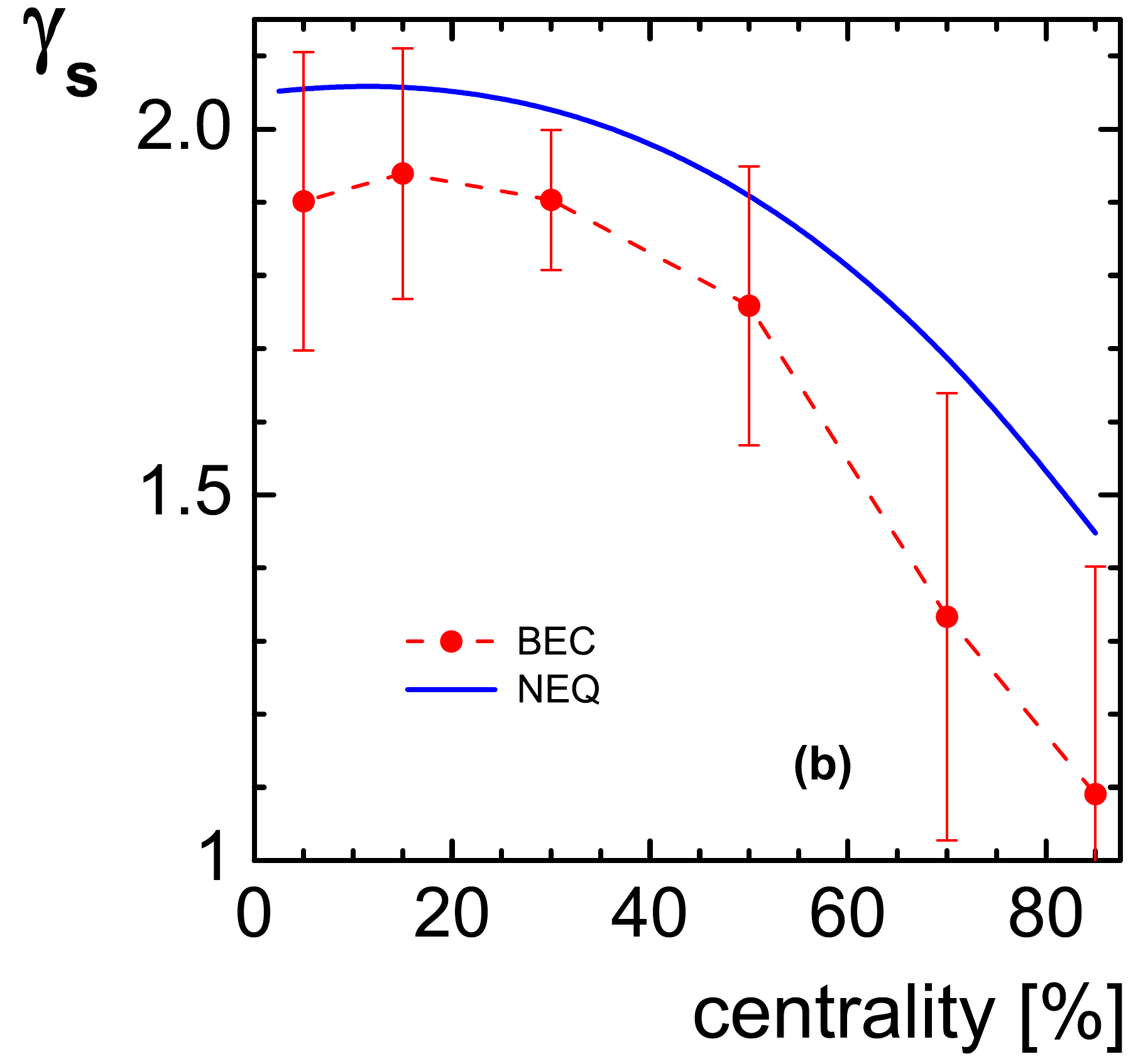,    width=0.49\textwidth}       
 \caption{(Color online) The non-equilibrium parameters $\gamma_q$ on the left and $\gamma_s$ on the right in BEC~\cite{Begun:2014aha} and NEQ~\cite{Begun:2014rsa}.
 The corresponding EQ values are equal to unity by definition.}\label{fig:gQ-gS}
\end{figure}

\begin{figure}[h!]
 \epsfig{file=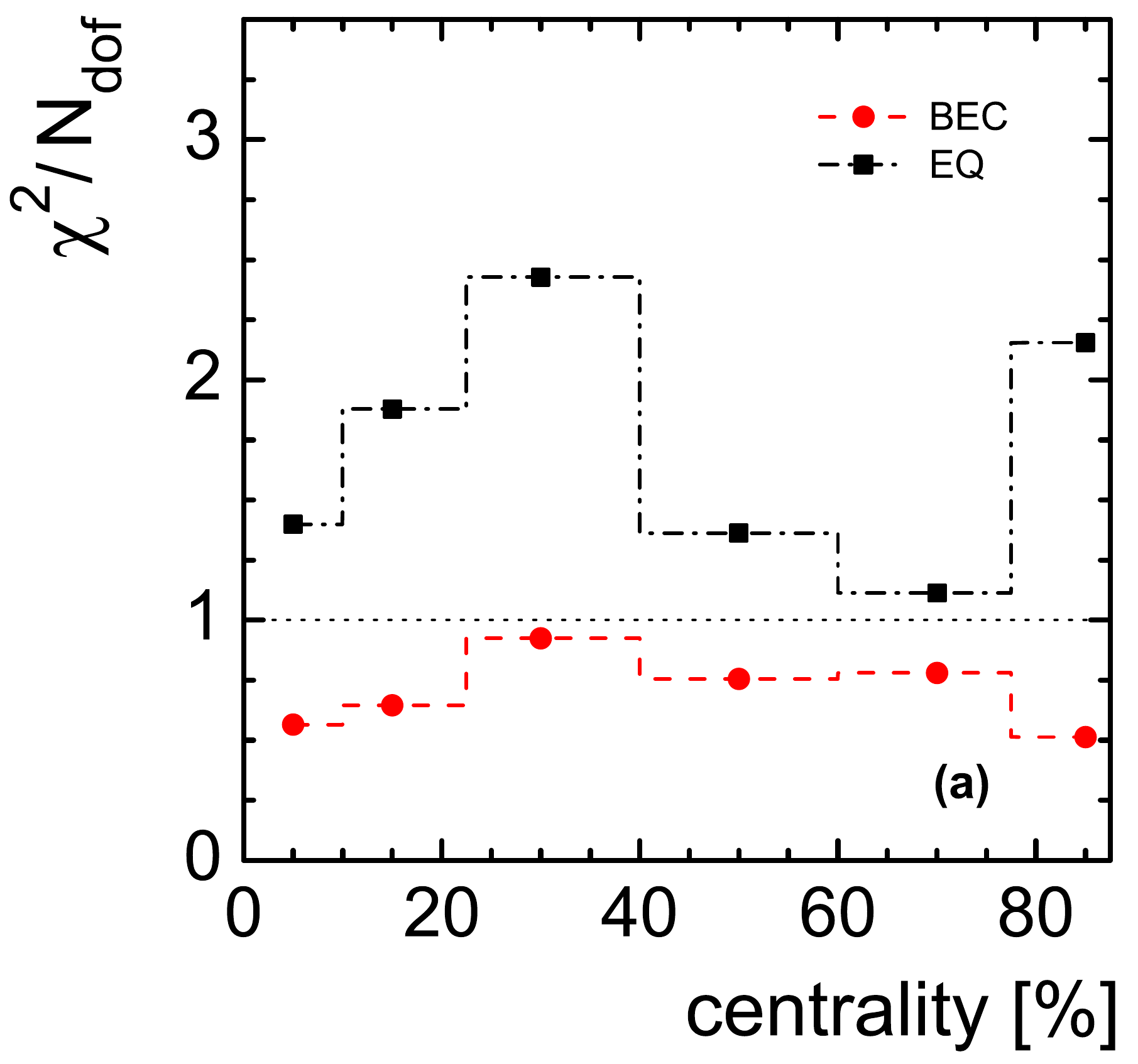,    width=0.49\textwidth}\;     
 \epsfig{file=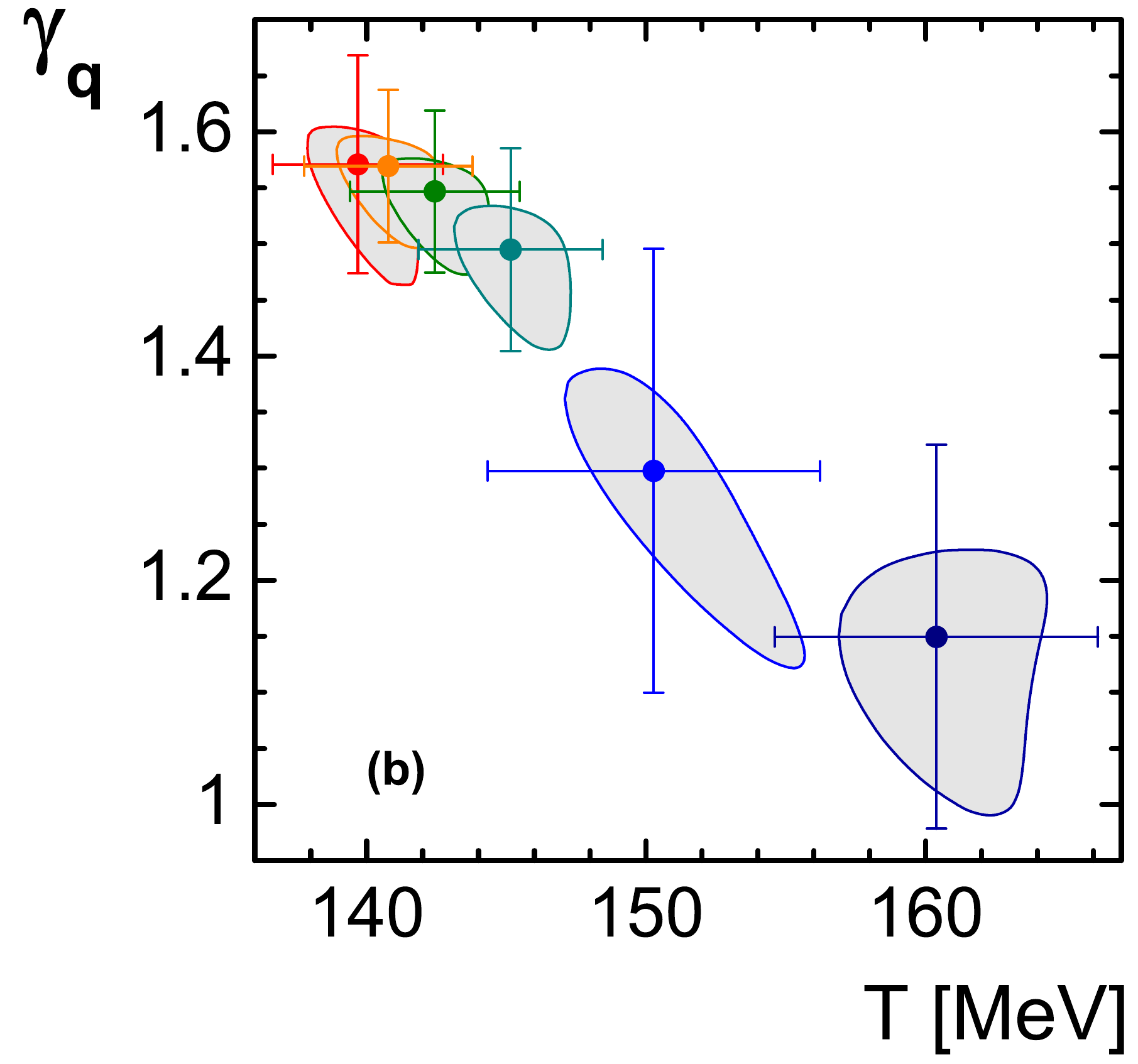,    width=0.49\textwidth}       
 \caption{(Color online) The $\chi^2/N_{\rm dof}$ in BEC and EQ on the left, and the best fit values for $\gamma_q$ and $T$ in BEC on the right together with the error bars.
 The contours on the right figure indicate the area where $\chi^2/N_{\rm dof}$ increases up to $10\%$.}\label{fig:Chi2-DataFit}
\end{figure}

Our results are shown in Figs.~\ref{fig:V-T},
\ref{fig:gQ-gS}~and~\ref{fig:Chi2-DataFit}.
The volume determined from the fits  is practically the same for
the NEQ and BEC models, because the number of particles
in the ground state appears to be relatively small. The
EQ volume is larger, because $\Upsilon>1$ makes
the system denser  in NEQ and BEC, compared to EQ.
Interestingly, the explicit treatment of the ground state
lowers the gap in temperature between equilibrium and
non-equilibrium models. These two versions of the model
even coincide for the peripheral collisions,
 as it was shown before in~\cite{Begun:2014aha}.

The left panel of Fig.~\ref{fig:Chi2-DataFit}  shows that the $\chi^2$
fit is much better for BEC than for EQ. Even the larger numbers of
degrees of freedom in EQ ($N_{\rm dof}$ = 12 for $c<80\%$ and
$N_{\rm dof}$ = 8 for $c= $~\ce{80}{90} in EQ, versus $10$ and $6$ in BEC,
respectively) do not help to decrease the $\chi^2/N_{\rm dof}$
values. The agreement with the data for individual particles is
also significantly better for BEC than for EQ. For example,
the  $K^*(892)^0$ resonance, which was not included in the fit,
deviates from the data by more than  $2\sigma$ in most central
collisions in EQ, similarly as in Ref.~\cite{Stachel:2013zma}, while
it agrees with the data within $1\sigma$ in BEC for each centrality. The
successful fit of the short living $K^*(892)^0$ with the parameters
obtained for long living particles is a strong argument in favor
of the single-freeze-out approximation.

The values of $\chi^2/N_{\rm dof}$ even drop below unity in BEC,
which means that the experimental errors are probably too large.
To check this behaviour, we calculated the $\chi^2/N_{\rm dof}$ contours and
found that the minimum of
$\chi^2/N_{\rm dof}$ is very shallow in BEC.
The right panel of Fig.~\ref{fig:Chi2-DataFit} shows the $10\%$ deviation from
the points with the best fit in the $T-\gamma_q$ plane for BEC.
The centrality increases from lower to higher temperatures, as in
Fig.~\ref{fig:V-T}. One may see that the increase of
$\chi^2/N_{\rm dof}$  by 10\% determines the area
that covers a substantial part of the region marked by
the error bars. This is so because there is a sudden rise in the values of
$\chi^2/N_{\rm dof}$ for increasing $\gamma_q$. This effect
elongates the contours towards smaller values of $\gamma_q$.
One may also notice an anti-correlation between $\gamma_q$ and $T$.

\begin{figure}[h!]
 \epsfig{file=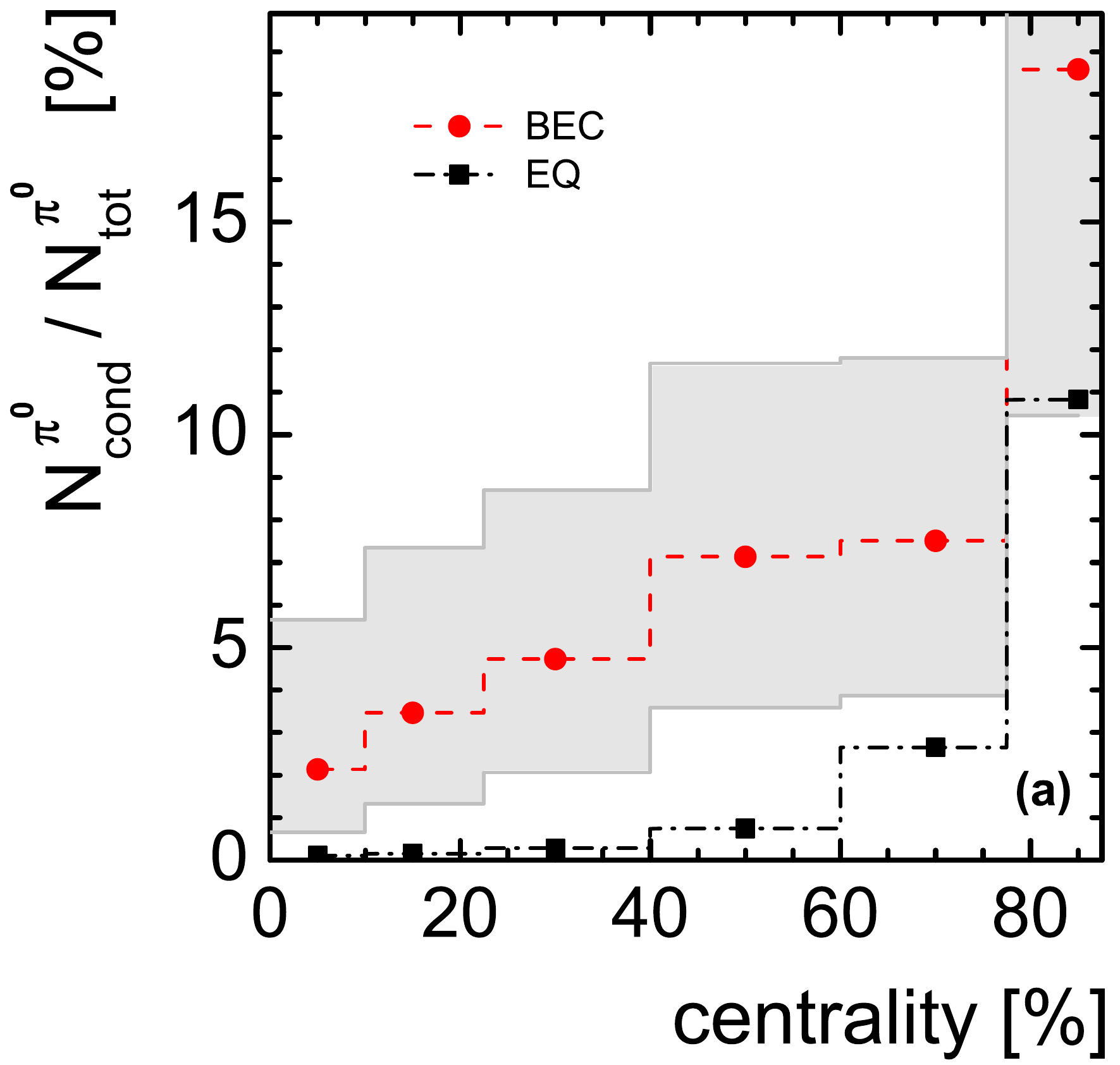,    width=0.49\textwidth}~      
 \epsfig{file=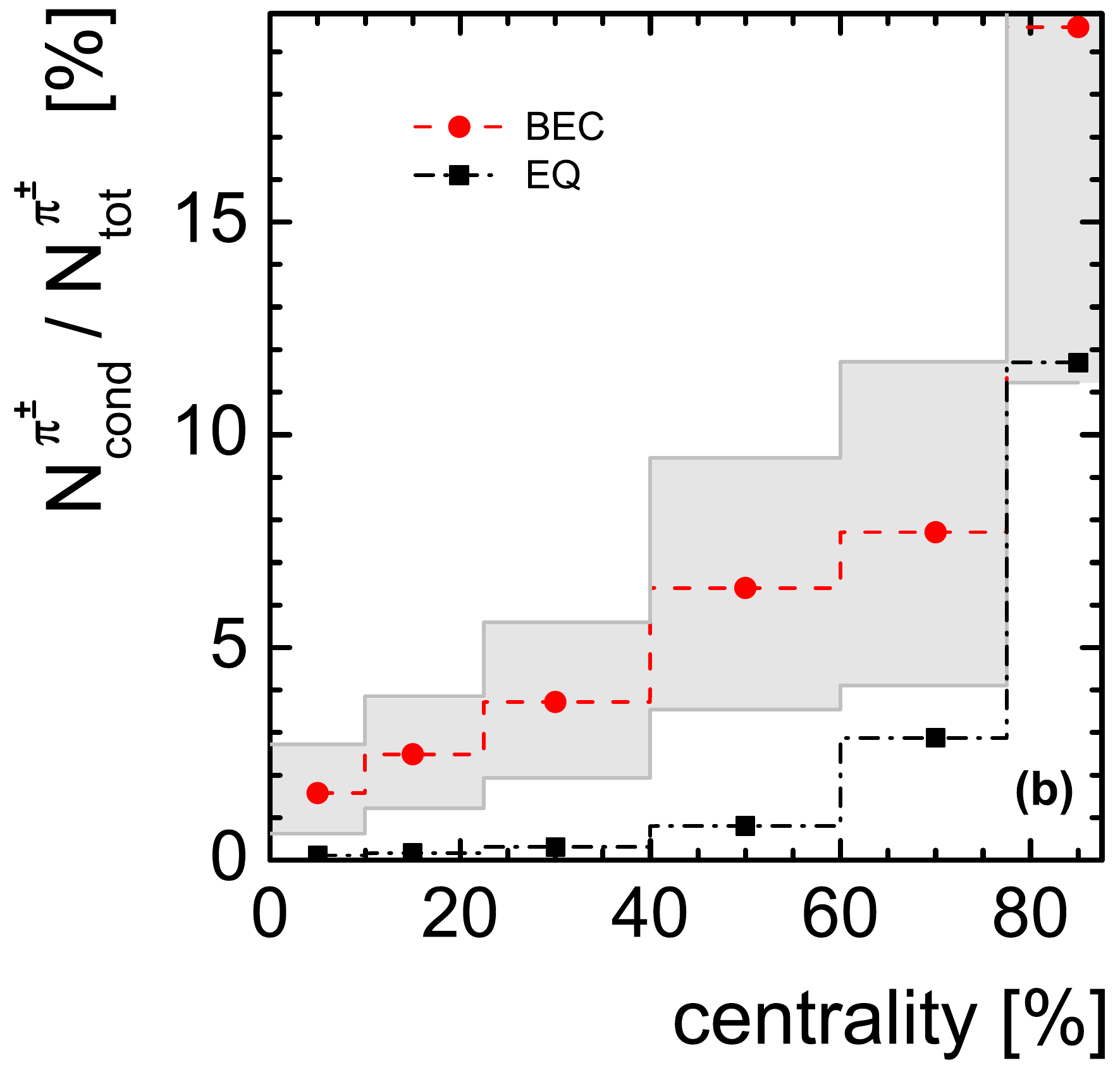,    width=0.49\textwidth}       
 \caption{(Color online) The ratio of neutral pions in the condensate on the ground state $N_{\rm cond}^{\pi^0}$ to the total number of neutral pions
 $N_{\rm tot}^{\pi^0}=N_{\rm cond}^{\pi^0}+N_{\rm norm}^{\pi^0}$ is shown on the left figure. The same for charged pions is shown on the right figure.
 The BEC lines were obtained using Eq.~(\ref{replace3}), while the EQ lines show the neglected contribution of the ground state in the EQ model,
 see Eq.~(\ref{Ncond-EQ}).}\label{fig:Rcond}
\end{figure}

The number of particles in the condensate in EQ, NEQ and BEC
models can be calculated using Eq.~(\ref{N0}) and our results
obtained with this formula are presented in  Fig.~\ref{fig:Rcond}. The grey
bands give the upper and lower bounds on the amount of condensate
in BEC. The width of the bands is determined by the
values of $T$ and $\gamma_q$ taken along the contours shown in
Fig.~\ref{fig:Chi2-DataFit} (other parameters are taken as the optimal).

The condensate rate in the NEQ model is not shown, as it is
very large. It means that using NEQ model, one should
always take into account the ground state effect. Otherwise, the obtained
$\gamma_q$ and $\gamma_s$ would be too high, while temperature
too small.

In BEC the condensate rate is $2\%$ in the most central collisions
and  $7-8\%$ in peripheral collisions (with a sudden increase to
$19\%$ in the most peripheral collisions). The plots for $\pi^0$
and $\pi^{\pm}$ are similar, because of the isospin symmetry that
we demand with the use of Eqs.~(\ref{Pi0}) and (\ref{Pi0Err}). The
upper bound is higher for $\pi^0$ in the most central collisions,
because temperature is lower there and the difference in masses of
$\pi^0$ and $\pi^{\pm}$ becomes more important.

The increase of the pion condensate contribution with centrality in the BEC model
is unexpected, because $\gamma_q$ and therefore $\mu_{\pi}$ decrease
with centrality. However, it has been observed already before \cite{Begun:2008hq} that
for smaller volumes the ratio of particles in the condensate increases
if the system's temperature is sufficiently high. Therefore,
the decrease of volume may produce the increase in the condensate
rate.

We also note that the ratio of the $\pi^0$ and $\pi^{\pm}$ number
in the zero momentum level (neglected in EQ calculations),
 \eq{\label{Ncond-EQ}
 N_{\rm cond}^{\rm EQ} ~=~
 \frac{1}{\exp\left(m_{\pi}/T_{\rm EQ}\right)-1}~,
 }
to the total number of $\pi^0$ and $\pi^{\pm}$-s also increases
for more than $10\%$. The number of particles in the ground state
$N_{\rm cond}^{\rm EQ}\simeq0.7$ in (\ref{Ncond-EQ}) for all
centralities, while $N_{\rm tot}^{\rm EQ}$ decreases from about
$700$ to $7$ at the most peripheral collisions,
see~\cite{Abelev:2013vea} for the $\pi^{\pm}$ data. Therefore the
ratio rises from $0.1\%$ to $10\%$. It means that the ground state
should be included for the analysis of very peripheral heavy-ion
or proton-proton collisions even in equilibrium models.

%
\section{Spectra}
\label{sect:BEC-THERMINATOR}
%

\subsection{Covariant Cooper-Frye formula with the condensate}

Equation (\ref{replace3}) can be formally obtained from
(\ref{V1}) by inserting the distribution function which contains a delta function
component
 \begin{equation}
 f({\bf{p}}) ~=~
 \frac{g}{\exp\left(\frac{\sqrt{{\bf p}^2+m^2}-\mu}{T}\right)-1}\,\left[~1~+~\frac{(2\pi)^3}{V}~\delta^{(3)}({\bf p})~\right]
 \label{fp}
 \end{equation}
in the integral over momentum (\ref{V1}).

In what follows we generalize Eq.~(\ref{fp}) to the form suitable for hydrodynamic and kinetic-theory applications. At first we note that the number of particles $N$ contained in the static box of volume $V$ may be obtained as the integral over the phase-space distribution function ${\bar f}(x,p)={\bar f}(t,{\bf x},{\bf p})$, namely
 \begin{equation}
 N ~= \int d^3x\,d^3p~{\bar f}(x,p).
 \label{Nbox}
 \end{equation}
Hence, for the particles in a static box  we may use the identification ${\bar f}(x,p)=(2\pi)^{-3}f({\bf{p}})$. If the particles are produced on the freeze-out hypersurface $\Sigma_\mu$, Eq.~(\ref{Nbox}) is generalised to the form
 \begin{equation}
 N ~= \int d\Sigma_{\mu} \int \frac{d^3p}{E_p}\,p^{\mu}~{\bar f}(x,p).
 \label{NSigma}
 \end{equation}
Then, the covariant generalisation of (\ref{fp}) for the systems which are boost-invariant and cylindrically symmetric is
\begin{equation}
 {\bar f}(x,p) ~=~ \frac{1}{(2\pi)^3} \frac{g}{\exp\left(\frac{p\cdot U -\mu}{T}\right)-1} \,\left[~1~+~\frac{(2\pi)^3}{\cal V}~
                   \delta(p \cdot X)~ \delta(p \cdot Y)~ \delta(p \cdot Z)~\right].
\end{equation}
Here ${\cal V} = \int d\Sigma_\mu u^\mu$ is the Lorentz invariant volume of the system, where the condensate is created. The quantity $U$ is the four-velocity of the fluid element at freeze-out,
while the four-vectors $X$, $Y$, and $Z$ define the three spatial
directions in the fluid local rest frame.  In the center-of-mass
system of the colliding nuclei, the form of  the four-vectors $X$,
$Y$, and $Z$ is~\cite{Florkowski:2010zz}:
 \begin{eqnarray}
 X ~&=&~(\sinh\theta_T\cosh\eta_{\parallel}~, \cosh\theta_T\cos\phi~, \cosh\theta_T\sin\phi~, \sinh\theta_T\sinh\eta_{\parallel} )~,\\
 Y ~&=&~(0~, -\sin\phi~, \cos\phi~, 0)~, \\
 Z ~&=&~( \sinh\eta_{\parallel}~, 0~, 0~, \cosh\eta_{\parallel} )~.
 \end{eqnarray}
Here $\theta_T$ is the transverse rapidity of the fluid element, $\phi$ is the azimuthal angle, and $\eta_\parallel$ is the space-time rapidity. In the local rest frame of the fluid element we have:  $X=(0,1,0,0)$, $Y=(0,0,1,0)$, $Z=(0,0,0,1)$. The projections of the four-momentum along $X$, $Y$, and $Z$ are
 \begin{eqnarray}
 p_X &=& p\cdot X =~ m_T\sinh\theta_T\cosh(y-\eta_{\parallel}) ~-~  p_T\cosh\theta_T\cos(\phi_p-\phi),    \\
 p_Y &=& p\cdot Y =~ p_T\sin(\phi-\phi_p), \\
 p_Z &=& p\cdot Z =~ m_T\sinh(\eta_{\parallel}-y).
 \label{pXpYpZ}
 \end{eqnarray}
Here we used the parameterisation of the four-momentum:  $p^{\mu}=(p^0,p^1,p^2,p^3)=(m_T\cosh
y,\,p_T\cos\phi_p\,,p_T\sin\phi_p\,,m_T\sinh y)$ where $p^0=E_p=\sqrt{{\bf p}^2+m^2}$ is the energy of a
particle, $y$ and $\phi_p$ are the particle rapidity and azimuthal
angle of the momentum, $p_T=\sqrt{(p^1)^2+(p^2)^2}$ is the
transverse momentum, and $m_T=\sqrt{p_T^2+m^2}$ -- the transverse
mass.

The Lorentz covariant integral in (\ref{NSigma}) is
\begin{equation}
 N^\mu ~= \int \frac{d^3p}{E_p}\,p^{\mu}~\frac{1}{(2\pi)^3} \frac{g}{\exp\left(\frac{p\cdot U -\mu}{T}\right)-1}\,
          \left[~1~+~\frac{(2\pi)^3}{\cal V}~\delta(p \cdot X)~ \delta(p\cdot Y)~ \delta(p \cdot Z)~\right].
\end{equation}
Because of the Lorentz covariance and quadratic dependence of the Bose-Einstein distribution of three-momentum, one can conclude that $N^\mu = n U^\mu$ with the density $n$ given by the expression
\begin{equation}
 n ~= \int d^3p\,~\frac{1}{(2\pi)^3} \frac{g}{\exp\left(\frac{\sqrt{{\bf p}^2 + m^2}-\mu}{T}\right)-1}\,
       \left[~1~+~\frac{(2\pi)^3}{\cal V}~\delta(p_x)~ \delta(p_y)~ \delta(p_z)~\right].
\end{equation}
In the Cracow model, the final expression for $\Delta N$ (the number of particles in the rapidity interval $\Delta y$) is
\begin{eqnarray}
\Delta N ~&=&  \int d\Sigma_\mu\, n \,  U^\mu =
\Delta y \, \pi r_{\rm max}^2 \tau_f \left( n_{\rm norm} + \frac{N_{\rm cond}}{\cal V}
\right)\nonumber \\
&\equiv&  \Delta y \frac{dV}{dy} \left( n_{\rm norm} + \frac{N_{\rm cond}}{\cal V}
\right).
\end{eqnarray}
Consequently, for the rapidity density one gets
\begin{eqnarray}
\frac{\Delta N}{ \Delta y } ~
= \frac{dV}{dy} \, n_{\rm norm} +  \frac{dV}{{\cal V}dy}  N_{\rm cond}.
\end{eqnarray}
In our case ${\cal V} =  \, \pi r_{\rm max}^2 \tau_f  \Delta y_{\rm cond}$, where $\Delta y_{\rm cond}$ is the rapidity range where the condensate may be formed. In this work we assume $\Delta y_{\rm cond}=1$. Consequently we get
\begin{eqnarray}
\frac{\Delta N}{ \Delta y } ~
= \frac{dV}{dy} \, n_{\rm norm} + N_{\rm cond}.
\end{eqnarray}
%

\subsection{The particle spectra with the local condensate}

The momentum distribution is obtained from the formula
 \eq{
 E_p\,\frac{dN}{d^3p} ~=~ \frac{dN}{dy\,d\phi_p\,p_T\,dp_T}
 ~= \int d\Sigma_{\mu}\,p^{\mu}~{\bar f}(x,p),
 }
where the condensate part of the distribution function can be written as
 \eq{\label{delta}
\delta(p_X)~\delta(p_Y)~\delta(p_Z)
~=~
 \frac{1}{p_T\,m^2}~\delta\left(\sinh\theta_T-\frac{p_T}{m}\right)~\delta(\phi-\phi_p)~\delta(\eta_{\parallel}-y).
 }
In our previous work we have found that the Cracow single freeze-out model can describe the LHC
data for $2.76$~TeV Pb+Pb collisions very well~\cite{Begun:2013nga,Begun:2014rsa}. Therefore,
at first we test effects of the Bose condensation using this model. In this case
$\sinh(\theta_T)=r/\tau_f$, where $r$ and $\tau_f$ are the
freeze-out radius and proper time, respectively. Using this relation we may further write
 \eq{\label{delta-Crc}
\delta(p_X)~\delta(p_Y)~\delta(p_Z)
~=~ \frac{1}{p_T\,m}~\delta\left(p_T-r\frac{m}{\tau_f}\right)~\delta(\phi-\phi_p)~\delta(\eta_{\parallel}-y).
 }
Using this result we obtain the final expression for the momentum distribution
 \eq{\label{d2Ndpt}
 \frac{dN}{dyd\phi_p\,p_Tdp_T}
 &~=~ \frac{g}{(2\pi)^3}
 \int_0^{2\pi}d\phi\int_{-\infty}^{\infty}d\eta_{\parallel}\int_0^{\infty}\theta(r_{\rm max}-r)\,rdr
 \nonumber
 \\
 &~\times~ \left[~ m_T\sqrt{\tau_f^2+r^2}\cosh(\eta_{\parallel}-y) ~-~ p_Tr\cos(\phi-\phi_p) ~\right]
 \nonumber
 \\
 &~\times~ \left\{\exp\left(\frac{\mu}{T}-\frac{1}{T}\left[~m_T\sqrt{1+\frac{r^2}{\tau_f^2}}\cosh(\eta_{\parallel}-y)~-~p_T\frac{r}{\tau_f}\cos(\phi-\phi_p)~\right]\right)
  ~\mp~1\right\}^{-1}
 \nonumber
 \\
 &~\times~ \left[~1~+~\frac{(2\pi)^3}{{\cal V}}
          ~\frac{\tau_f}{p_T\,m^2}~\delta\left(r-p_T\frac{\tau_f}{m}\right)~\delta(\phi-\phi_p)~\delta(\eta_{\parallel}-y)~\right]~,
 }
where $\theta(r_{\rm max}-r)$ is the Heaviside step function.

The first term in the (\ref{d2Ndpt}) is the usual one, while the
second term with the delta function corresponds to the Bose
condensate. The integrals over $\phi$ and $\eta_{\|}$ are
cancelled by the delta functions and we obtain:
 \begin{eqnarray}
 \label{dNcond}
 \frac{dN}{dyd\phi_p\,p_Tdp_T}\Bigg|_{\rm cond} &=&  \frac{g}{{\cal V}}
 \int_0^{\infty} \theta(r_{\rm max}-r) rdr
 \left[~ m_T\sqrt{\tau_f^2+r^2} ~-~ p_Tr ~\right]~\frac{\tau_f}{p_T\,m^2}~\delta\left(r-p_T\frac{\tau_f}{m}\right)
  \nonumber  \\
 &&~\times~\left\{\exp\left(\frac{\mu}{T}-\frac{1}{T}\left[~m_T\sqrt{1+\frac{r^2}{\tau_f^2}}~-~p_T\frac{r}{\tau_f}~\right]\right)~\mp~1\right\}^{-1}  \\
 &&~=~ \frac{g}{{\cal V}}\,\frac{\tau_f^3}{m^2}~ \theta \left(r_{\rm max} - p_T \tau_f/m \right)
 \left\{\exp\left(\frac{m-\mu}{T}\right)~\mp~1\right\}^{-1}
 \nonumber \\
&& =~ \frac{1}{{\cal V}}\,\frac{\tau_f^3}{m^2} \theta \left(r_{\rm max} - p_T \tau_f/m \right) \,\,N_{\rm cond} ,
 \nonumber
 \end{eqnarray}
where $N_{\rm cond}$ is the number of particles in Bose condensate
as in (\ref{replace3}). It follows from equation~(\ref{dNcond})
that Bose condensate adds a constant number of particles to the
usual momentum spectrum and these particles may have a momentum up
to
 \eq{\label{ptmax}
 p_T^{\rm max}~=~m\,\frac{r_{\rm max}}{\tau_f}~.
 }

\begin{figure}[h!]
 \epsfig{file=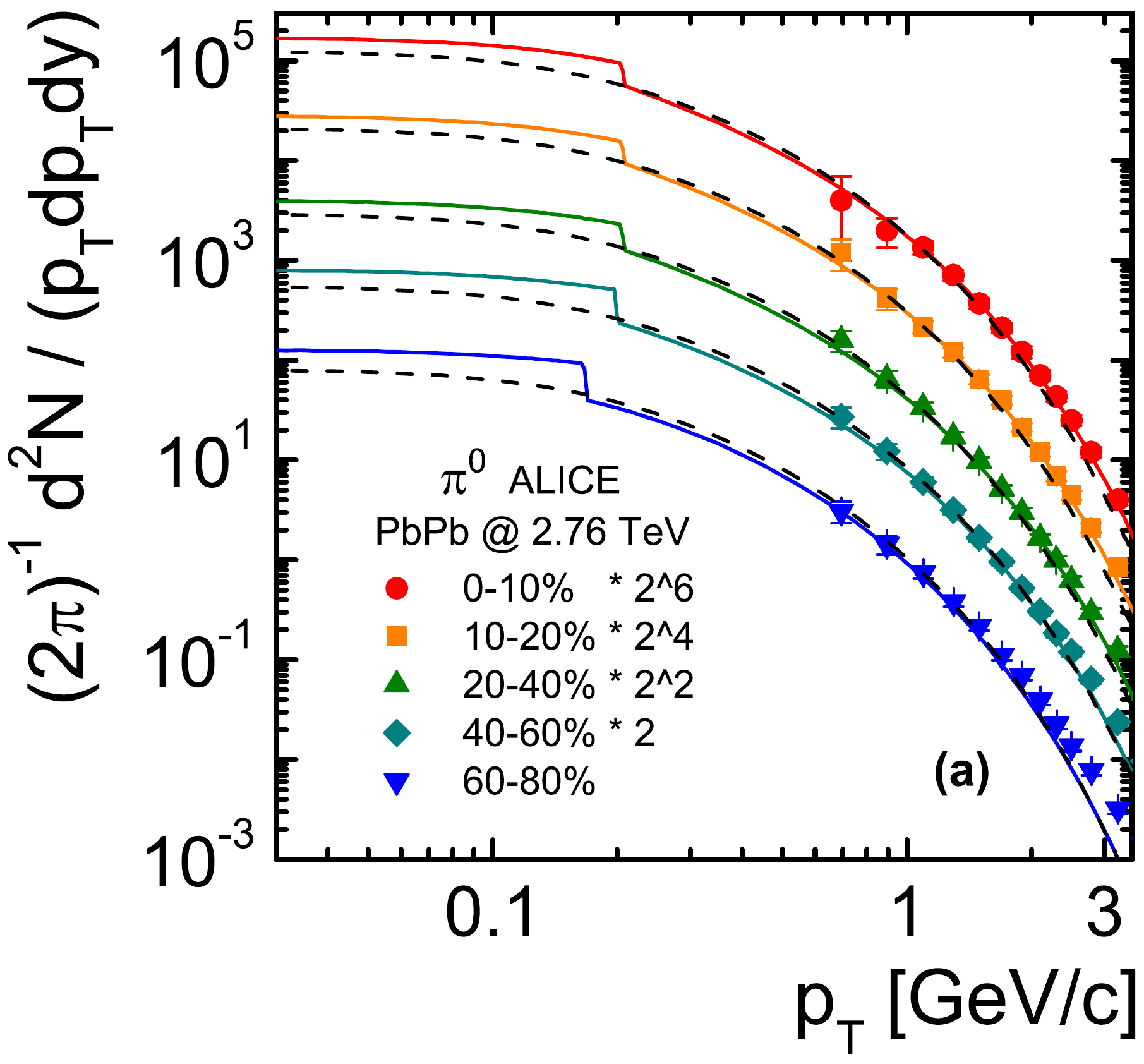,    width=0.49\textwidth}\;     
 \epsfig{file=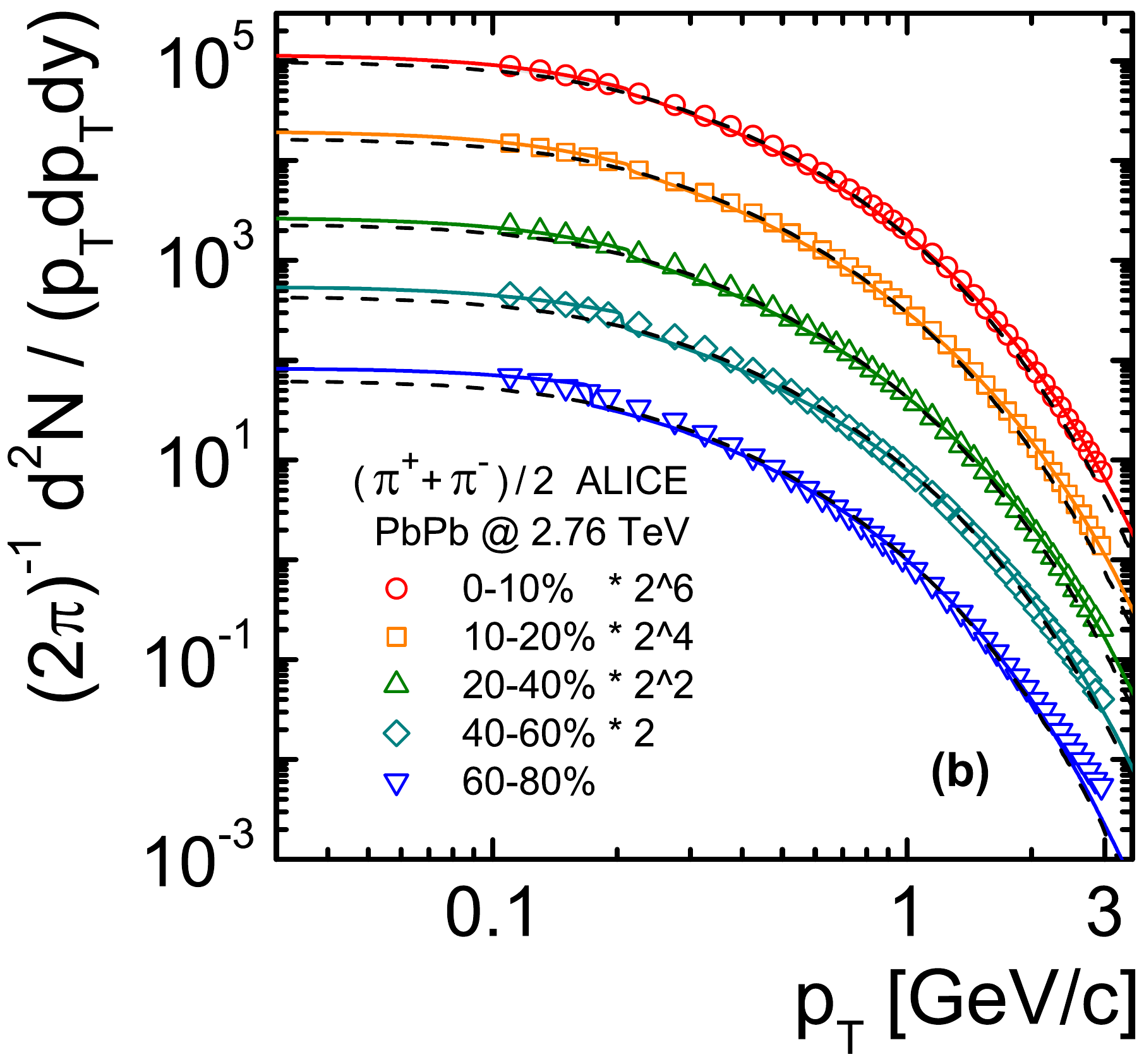,    width=0.49\textwidth}       
 \caption{(Color online) The $p_T$ spectra calculated in the BEC model  (solid lines) and EQ model (dashed lines) at different centralities
 are compared with the existing data  (symbols) for neutral pions~\cite{Abelev:2014ypa} on the left and with the data for charged
pions~\cite{Abelev:2013vea} on the right.}\label{fig:d2Nptdptdy}
\end{figure}

We added the zero momentum level into THERMINATOR using the
Dirac delta function (\ref{delta-Crc}). It turns out that in
order to reproduce the spectra we may keep the same
$r_{\rm max}/\tau_f$ ratio as in our previous
paper~\cite{Begun:2014rsa} and just rescale $r_{\rm max}$ and
$\tau_f$ in such a way that we obtain the volume
determined from the studies of multiplicities described in
Sec.~\ref{sect:Ratios}. The results for pions are shown in
Fig.~\ref{fig:d2Nptdptdy}. The error bars for
$\pi^{\pm}$ are not shown, because they are of the
order of the symbol size.

One may notice a very good overall agreement. In the BEC
model the spectra of other particles including protons, kaons,
$K^*(892)^0$ and $\phi(1020)$, also agree with data, and are
almost identical to those that we obtained
in~\cite{Begun:2014rsa}.
The agreement with data for pions at high $p_T$ is even improved
compared to~\cite{Begun:2014rsa}, because we use the same $r_{\rm
max}/\tau_f$, but have bigger temperature in BEC than in NEQ.
Therefore there are more high-$p_T$ particles in BEC than in NEQ.

On the other hand, the agreement of pions in the EQ model with the
data is rather good, but other particles are described much
worse~\cite{Begun:2014rsa}. For example, protons in EQ model
deviate from data for more than the factor of two in the most
central collisions at low $p_T$~\cite{Begun:2014aha}.

\begin{figure}[h!]
 \epsfig{file=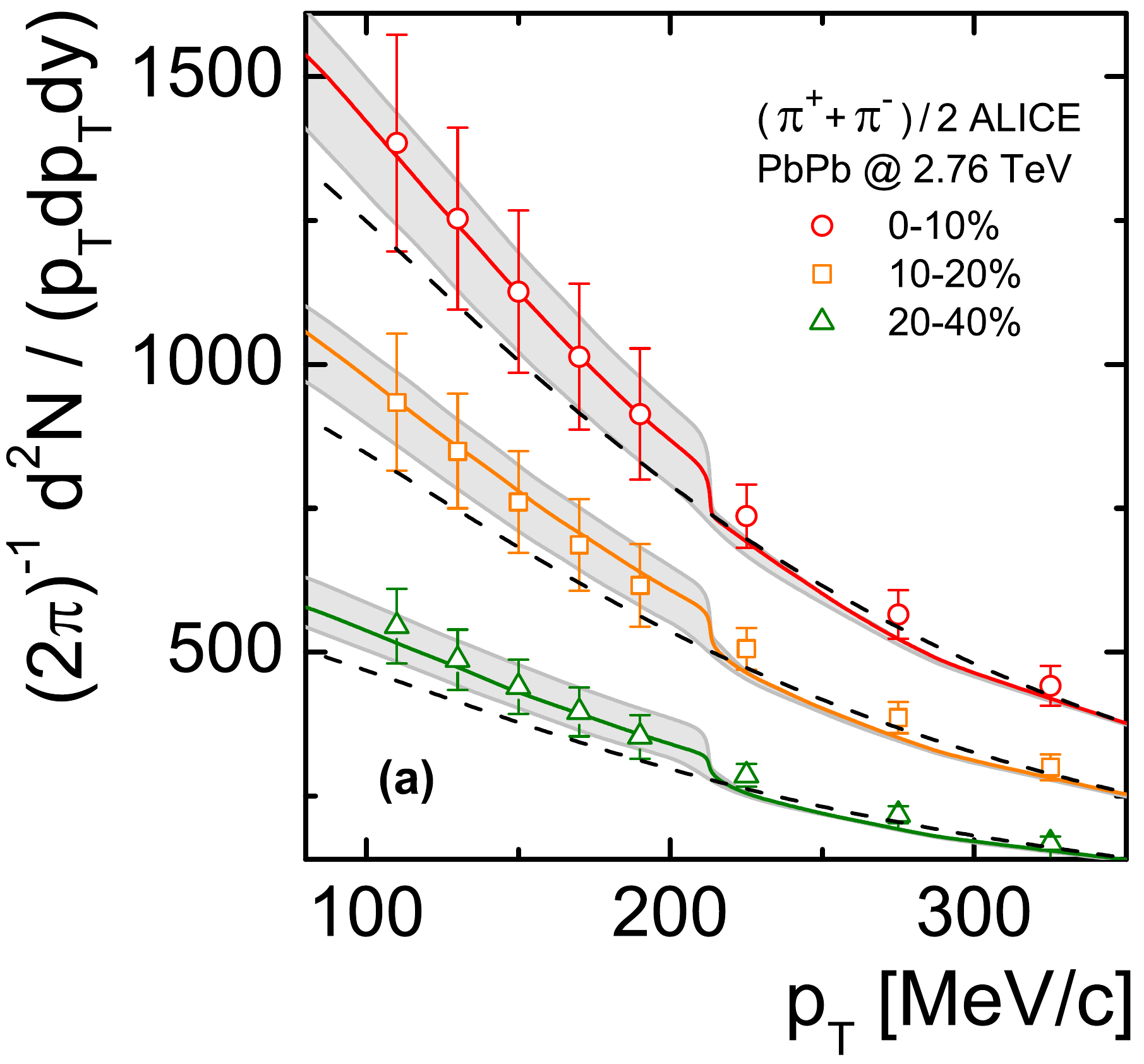,    width=0.49\textwidth}~      
 \epsfig{file=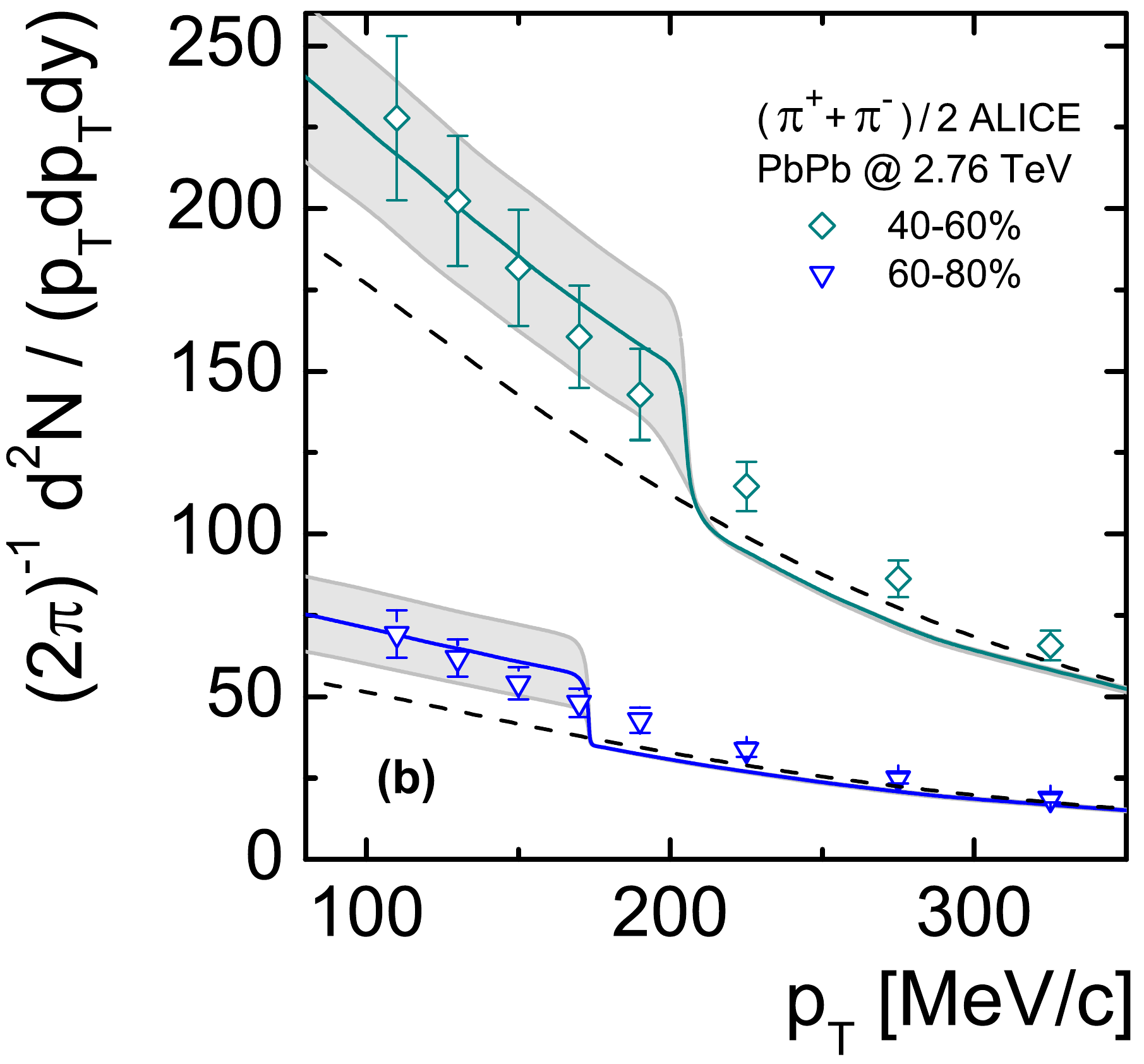,    width=0.49\textwidth}       
 \caption{(Color online) The low momentum part of the spectra for charged pions in central and semi-central collisions on the left and in
 peripheral collisions on the right are compared to the calculations in the BEC  (solid lines) and EQ (dashed lines) models. The grey bands correspond to minimal
 and maximal condensate rate shown in Fig.~\ref{fig:Rcond}, which is obtained assuming $10\%$ increase of the $\chi^2/N_{\rm dof}$,
 see Fig.~\ref{fig:Chi2-DataFit} right.
 }\label{fig:PtLow}
\end{figure}

The largest difference for pions is also seen at low $p_T$. There
is a step at $p_T = p_T^{\rm max}$  in the BEC model, see
Eq.~(\ref{ptmax}), because of the contribution of the particles
from the ground state. This is the artefact of the approximation
that all particles from the low discrete levels are accumulated in
the ground state~(\ref{replace3}).
The step is larger for neutral pions, because they are lighter and
therefore condense earlier. Unfortunately, $\pi^0$'s have not been
measured yet at low $p_T$. There are only several points for
charged pions, $\pi^{\pm}$, which are plotted in the linear scale
without the $2^n$ factors in Fig.~\ref{fig:PtLow}.
One can see that the EQ lines do not describe the data in the
low $p_T$ region for the very peripheral collisions. For the most
central collisions the EQ lines go along lower edges of the error
bars. This cannot be improved, because the EQ model overshoots protons
at these centralities~\cite{Begun:2014rsa,Begun:2014aha}.
Therefore it would be impossible to describe in EQ neither
low $p_T$ pions nor protons when the experiment  provides the
improved data with lower error bars.

The BEC line with the step is within the error bars for the most
central collisions, but the step is too big for more peripheral
collisions. However, the smallest $p_T$ points are fitted at all
centralities. One may also notice that the amount of extra
particles to the left of the step is approximately equal to the
number of missing particles to the right of the step. Therefore
the inclusion of several more levels would lead to finer steps and
smoother line, which goes within the error bars.

The grey bands indicate the maximal and minimal amount of
condensate at the $10\%$ higher $\chi^2/N_{\rm dof}$ from the best
fit, see Fig.~\ref{fig:Chi2-DataFit} right and
Fig.~\ref{fig:Rcond}. The grey band is within the error bars for
most central collisions. One can even double the contribution from
condensate there. However, in the most peripheral collisions the
upper bound for the condensate is definitely too high and should
be decreased by a factor of two\footnote{One should remember that
the discussed low-$p_T$ region of $\pi^{\pm}$ spectra gives a
small contribution to the mean multiplicities, therefore the
spectra are much more sensitive to the condensate.}. Thus, the
combined data on mean multiplicities and spectra are compatible
even with the constant condensate rate at the level of $5\%$.

%
\section{Conclusions}
\label{sect:CONCLUSIONS}
%

In this work we have analysed the possibility of the pion condensation
in heavy-ion collisions at the LHC energies. This phenomenon has attracted
a lot of attention in the past (see \cite{Shuryak:2014zxa} and
references therein).

Our study shows that the pion condensate rate in the present data
is $2\%$ in the most central collisions
and  $7-8\%$ in peripheral collisions (with a sudden increase to
$19\%$ in the most peripheral collisions). The obtained $\chi^2/N_{\rm dof}$
values for the thermodynamic parameters
have a shallow minimum. It results in a large uncertainty in the
determination of the exact number of pions in the condensate.
The spectra show an opposite trend with centrality
allowing larger condensate contributions in the most
central collisions as compared to the peripheral ones.

The data for charged pion spectrum at the \ce{80}{90} centrality
would help to confirm  the  increase of the
condensate rate with centrality. However, the most important are
the measurement of the $\pi^0$ multiplicities and the $\pi^0$
spectra in the low $p_T$ region.

\acknowledgments

We thank Mateusz Ploskon for sending us the data on $\pi^0$
spectra. V.B and W.F. were supported by Polish National Science
Center grant No. DEC-2012/06/A/ST2/00390.


\end{document}